\documentclass{ws-ijmpd_pol}

\begin{document}

\markboth{P. Bordas, V. Bosch-Ramon and J. M. Paredes}
{Gamma-rays from SS~433 and its interaction with the W50 nebula}

%
\catchline{}{}{}{}{}
%

\title{Gamma-rays from SS~433 and its interaction with the W50 nebula}

\author{Pol Bordas}

\address{ Departament d'Astronomia i Meteorologia and Institut de Ci\`encies del
Cosmos (ICC), Universitat de Barcelona (UB/IEEC), Mart\'{i} i Franqu\`{e}s 1, E08028 Barcelona, Spain \\
pbordas@am.ub.es}

\author{Valent\'i Bosch-Ramon}

\address{Max Planck Institut f\"ur Kernphysik, Saupfercheckweg 1, D69117
Heidelberg, Germany\\
vbosch@mpi-hd.mpg.de}

\author{Josep Maria Paredes}

\address{ Departament d'Astronomia i Meteorologia and Institut de Ci\`encies del
Cosmos (ICC), Universitat de Barcelona (UB/IEEC), Mart\'{i} i Franqu\`{e}s 1, E08028 Barcelona, Spain \\
jmparedes@ub.edu}

\maketitle


\begin{abstract}
We investigate the production of gamma-rays in the inner regions of SS 433 and in its interaction
between its jets and the W50 nebula. We estimate the VHE emission that can be generated within
the jets at distances $\gtrsim 10^{13}$~cm from the compact object. We also apply a theoretical model of
the jet/medium interaction to SS 433/W50. We compare the predicted fluxes to
observations of SS 433 at TeV energies, and derive new constraints of the physical properties of
this system. 
\end{abstract}


\section{Introduction}	

Relativistic jets from a stellar compact object were discovered for the first time in the microquasar SS~433 \cite{Spencer1979,Abell1979}. The system is composed of a 9~M$_{\odot}$ black hole orbiting a 30~M$_{\odot}$ A3--7 supergiant star with orbital radius $\sim$~79~$R_{\odot}$ \cite{Fabrika2004,Cherepashchuk2005} and period $P_{\rm orb} \sim 13.1$~d. Located at a distance of 5.5~$\pm$~0.2 kpc \cite{Blundell2004}, the system shows relativistic jets with a velocity of 0.26$c$ \cite{Eikenberry2001}. The jets precess with a period $P_{\rm pre} \sim 162$~d in cones of half opening angle $\theta \approx$ 21$^{\circ}$ and are inclined by $i\approx$ 78$^{\circ}$ with respect to the observer line of sight.\cite{Eikenberry2001}. SS~433 is the only X-ray binary system in which hadrons have been found in the jet. Clouds of plasma with baryonic content are observed at large distances, suggesting that atomic reheating is working at $z_{\rm jet}\lesssim 10^{17}$~cm from the compact object. Furthermore, a continuous regime of supercritical accretion onto the black hole is accomplished in SS~433. This could explain the large kinetic power thought to be transported by the jets, $L_{\rm k}\sim$~10$^{39}$~erg~s$^{-1}$ (see, e.g. Ref.~\refcite{Dubner1998}).


\begin{figure}[pb]
\centerline{\psfig{file=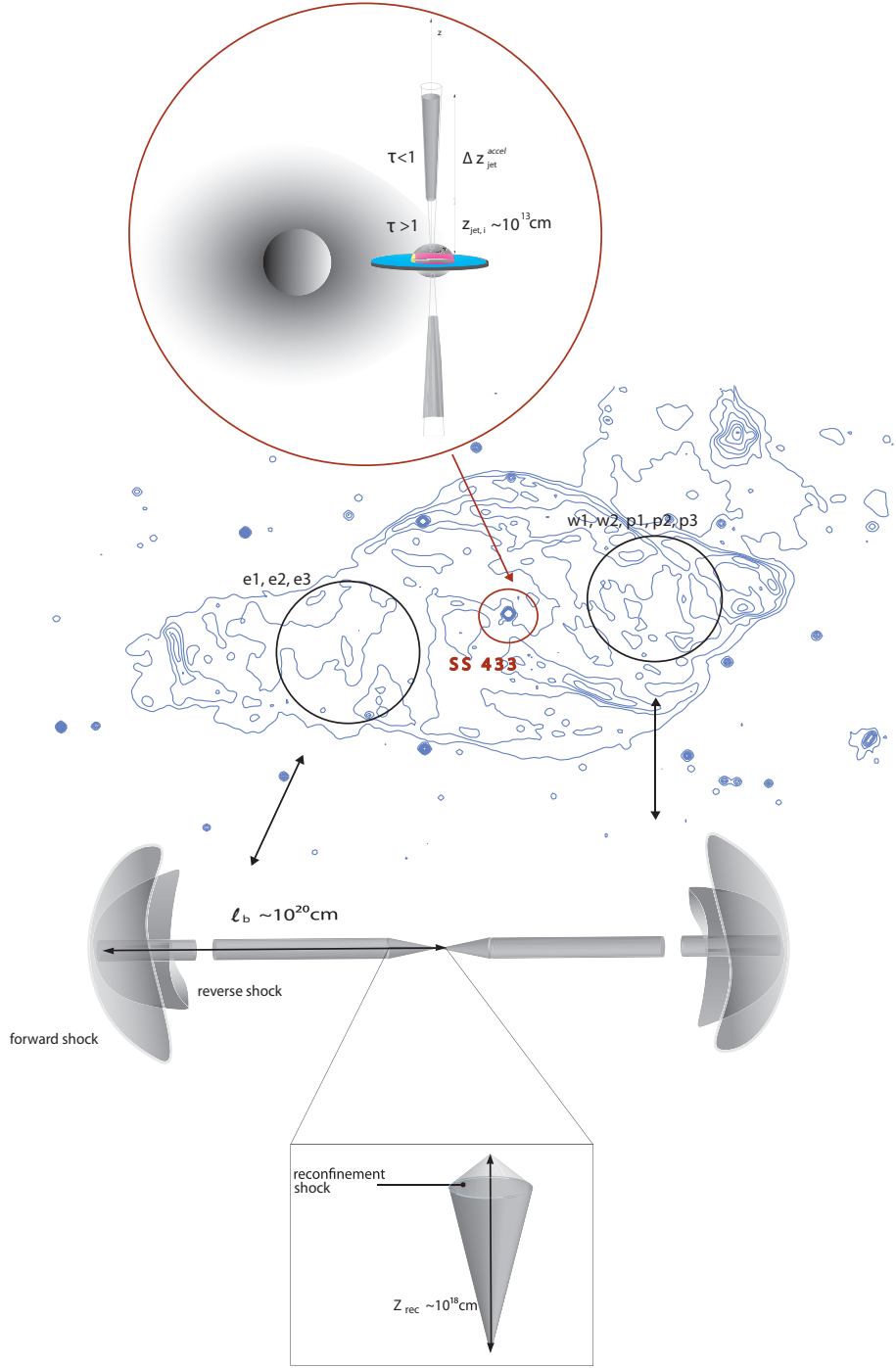,width=12.cm}}
\vspace*{8pt}
\caption{Radio map of the extended W50 radio nebula (blue contours, extracted from Dubner et al. 1998) and the different regions where particle acceleration and emission could take place. \textbf{Top}: sketch of the SS~433 central engine. The jet acceleration region $\Delta z$ where the optical depth is $\le~1$ is represented by gray-shadowed cones. See text for details. \textbf{Bottom}: East and West interaction regions. The forward, reverse and reconfinement strong shocks are displayed. See Bordas et al. (2009) for a complete description of the formation of the interaction structures.\label{f1}}
\end{figure}


\section{VHE observations of SS~433/W50}	

SS~433/W50 has been extensively observed by the HEGRA\cite{Aharonian2005}, MAGIC\cite{Saito2009} and CANGAROO-II\cite{Hayashi2009} Cherenkov telescopes. No gamma-ray signal has been found, and upper limits at different energy thresholds have been established both for the inner system and the different interaction regions (the e1, e2, and e3 regions at the east and the w1, w2, p1, p2, and p3 at the west; see Refs.~\refcite{Aharonian2005}, \refcite{Saito2009} and \refcite{Hayashi2009} for details). In the table below we list some of the reported upper limits. MAGIC observations, in particular, took into account the strong absorption due to the periodic companion eclipses as well as the attenuation due to the precession of the accretion disk envelope (see Ref.~\refcite{Reynoso2008} for a detailed study).


\begin{table}[ph]
\tbl{Observations of SS~433 and the jet/W50 interactions in the TeV regime by the MAGIC, H.E.S.S. and CANGAROO-II Cherenkov telescopes. The total observation time, the energy threshold $E_{\rm th}$ and the reported upper limits to the integral flux $\phi_{\rm ul}$ (in units of $ 10^{-12}$~ph~cm$^{-2}$~s$^{-1}$) are listed.}
{\begin{tabular}{@{}|c|cccc|c|cccc|@{}} 
\hline
Source & IACT & obs. time (h) & $E_{\rm th}$ (GeV) &$\phi_{\rm ul}$ & Source & IACT & obs. time (h) & $E_{\rm th}$ (GeV) &$\phi_{\rm ul}$  \\ \hline
SS~433  & HEGRA & 96.3 & 800 & 8.93 & p1      & CANGAROO-II & 85.2 & 850 & 1.5\\
        & MAGIC & 6.6 & 250 & 4.30 & p2      & CANGAROO-II & 85.2 & 850 & 1.3\\
e1      & HEGRA & 72.0 & 800 & 6.18 & p3      & CANGAROO-II & 85.2 & 850 & 0.79\\
e2      & HEGRA & 73.1 & 800 & 9.18 & w1      & HEGRA & 104.9 & 800 & 6.65\\
e3      & HEGRA & 68.8 & 800 & 8.96 & w2      & HEGRA & 100.7 & 800 & 9.00\\ 
\hline
\end{tabular} \label{ta1}}
\end{table}


\section{Gamma-ray production in SS~433/W50}

\subsection{Inner regions}

We have estimated the Inverse Compton (IC) VHE fluxes produced at the borders of the binary system (see the upper inset in Fig.~1). At distances $z_{\rm jet}\le 10^{13}$~cm along the jet, the opacity to gamma-ray propagation due to the companion and accretion disk photon fields should make the optical depth $\tau_{\gamma} > 1$. We consider therefore acceleration/emission regions beginning further away than that point. The free parameters in our model are the non-thermal acceleration fraction $q_{\rm rel}$ and the accelerator/emitter size $\Delta z = z_{\rm jet, f}-z_{\rm jet, i}$. Table~2 lists some of the parameters that remain fixed in our model and their respective values (parameters from the interaction model are also displayed, see below).

We assume the dominance of the IC channel over any other relevant energy loss mechanisms. To estimate the $\gamma$-ray fluxes we approximate both the companion star and the disk envelope as point-like sources of isotropic black-body radiation at temperatures $T$=8500 K and 21.000 K, respectively \cite{Gies2002,Fuchs2006}. We use the IC cross section including both Thomson and Klein-Nishina regimes for this process, since $\gamma_{\rm e} \gtrsim \gamma_{\rm KN} \equiv (4\epsilon_{0})^{-1}$, where $\gamma_{\rm e}$ is the electron Lorentz factor and $\epsilon_{0} \times m_{\rm e}c^{2}\sim 2.7 \, K_{\rm B}T$ is the peak energy of each corresponding photon field. We take a given fraction of the total jet power to be delivered to a leptonic accelerated plasma that follows a power-law distribution, $N_{\rm e} \propto \gamma_{\rm e}^{-p}$, with a spectral index $p = 2$. A maximum particle Lorentz factor $\gamma_{\rm e}^{max} = 10^{6}$ is used in our computations.

%
%

\subsection{Interaction regions}

Three strong shocks are formed when the jet impacts with the surrounding nebula (see Fig. 1, bottom). A forward shock, propagating into the medium and enclosing the shell region; a reverse shock directed into the jet material and inflating the cocoon, and a recollimation shock that appears when the pressure in the conical jet equals that of the cocoon. We have studied the non-thermal emission coming from these three regions through synchrotron, IC and relativistic bremsstrahlung (the last only in the shell, where it is relevant) emission channels. The model assumes a self-similar growing of the interaction structures to estimate the physical properties of each shocked region. The reader is referred to the work presented in Ref.~\refcite{Bordas2009} for a complete model description. Some relevant model parameters are listed in Table~2.


\begin{table}[ph]
\tbl{List of the parameters that remain with a constant value in the analytical model.}
{\begin{tabular}{@{}lcc@{}} \toprule
\it{Parameter} & \it{Symbol} & \it{Value} \\  \colrule  

Jet kinetic power (erg~s~$^{-1}$)& $Q_{\rm jet}$ & $10^{39}$ \\ 
ISM density (cm$^{-3}$)  & $n_{\rm ISM}$ & 1  \\ 
Source age (yr) & $t_{\rm MQ}$  & $5\times 10^{4}$ \\ 
Jet Lorentz factor  & $\Gamma_{\rm jet}$ & $1.04$ \\ 
Jet semi-opening angle ($^\circ$)  & $ \Psi $ &  1.2   \\ 
Luminosity companion star (erg~s~$^{-1}$)  & $L_{\star}$ & $10^{39}$ \\ 
Self-Similar parameter  & $ R $ & $3$     \\ 
Magnetic equipartition fraction & $ \eta $ & $0.1$   \\  \botrule
\end{tabular} \label{ta1}}
\end{table}


\section{Model results}

Different integral luminosities from the inner system regions for energies $E_{\gamma} \ge 250$~GeV are obtained if gamma-rays are generated at distances $z_{\rm jet,i} \sim 10^{13-14}$~cm up to $z_{\rm jet,f} \sim 10^{17}$~cm. We note that taking $z_{\rm jet,f}$ much higher than $10^{17}$~cm does not significantly change our results, since at these distances the photon field energy density is negligible (we do not consider the contribution from the CMB in our calculations). The integral fluxes produced along the accelerator/emitter region $\Delta z_{\rm jet}^{accel}$ as a function of the injection point and the acceleration efficiency are shown in Fig.~2. The MAGIC upper limit $\phi_{\rm u.l} \le 4.3 \times 10^{-12}$ ph cm$^{-2}$ s$^{-1}$ for $E_{\gamma} \ge 250$ GeV for the central regions of SS~433 \cite{Saito2009} is also displayed.



\begin{figure}[t]
\centerline{\psfig{file=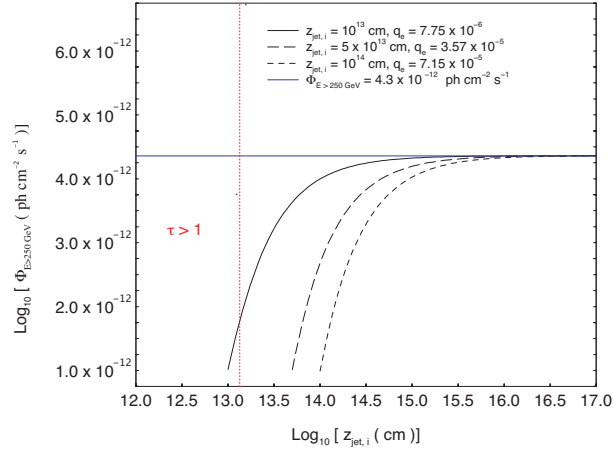,width=8.cm}}
\vspace*{8pt}
\caption{MAGIC integral flux upper limit at E$_{\gamma}\ge$~250 GeV (horizontal blue line), together with the total flux obtained through IC process for three different values of the acceleration efficiency $q_{\rm rel}$ and acceleration injection height $z_{\rm jet, i}$ (we recall that leptons are accelerated along $\Delta z_{\rm jet}^{accel} = z_{\rm jet,f}-z_{\rm jet,i}$, and we have used  $z_{\rm jet,f} = 10^{17}$~cm in all cases; see text for details). The region at the left side of the red-dotted line has optical depth values $\ge 1$. \label{f2}}
\end{figure}



\begin{figure}[h]
\centerline{\psfig{file=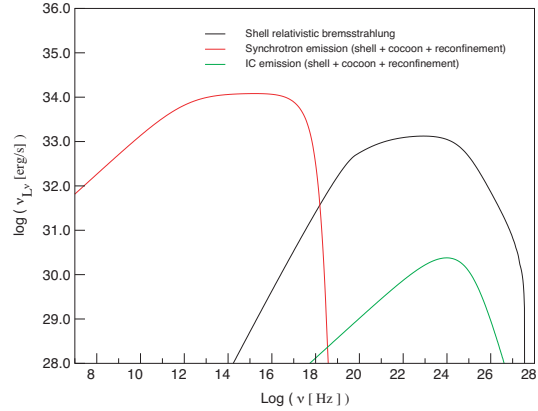,width=7.cm}}
\vspace*{8pt}
\caption{SED of the interaction regions obtained for the parameters listed in Table~2. The sum of the contributions from the shell region, the cocoon and the reconfinement region is displayed. Relativistic bremsstrahlung is important only in the shell region, due to the low particle densities in the jet and cocoon. The contribution of only one jet impacting on the nebula is showed. \label{f3}}
\end{figure}


Concerning the interaction of SS~433 with the W50 nebula, we show in Fig.~3 the SED computed for the shell, the cocoon and the reconfinement interaction regions. Relativistic bremsstrahlung is the most efficient gamma-ray production mechanism in the shell zone, reaching bolometric luminosities at a level $ \sim 10^{33}$~erg~s$^{-1}$. IC is the main radiation channel at high and VHE in the cocoon and reconfinement regions, with bolometric luminosities at the level of a few~$\times 10^{31}$~erg~s$^{-1}$. The integral gamma-ray flux for energies $E \ge 800$~GeV computed from the SED is found to be slightly lower than the upper limits for the different interaction regions listed in Table~2.

\section{Discussion}

We explored the $\gamma$-ray emission produced by relativistic electrons both in the inner jet regions and in the jet/W50 interaction regions. The gamma-ray fluxes obtained for the central regions depend linearly on $q_{\rm rel}$ and the acceleration region within the jet, which are nonetheless poorly constrained under a theoretical point of view. A value of $q_{\rm rel}$ far larger than $\sim 10^{-6}$ for $\Delta z_{\rm jet}^{accel}$ starting at $z_{\rm jet,i} \sim 10^{13}$~cm is ruled out within our flux estimations and the reported upper limits, while $q_{\rm rel} \ge 10^{-5}$ is not allowed provided that the acceleration region $\Delta z_{\rm jet}^{accel}$ starts at $z_{\rm jet,i} \ge 10^{14}$~cm. Acceleration processes could be much less efficient than expected, though it seems unlikely since acceleration is actually required to explain the non-thermal emission at lower energies. In addition to the IC process, relativistic bremsstrahlung and SSC could be important at the very inner regions near the jet base, where magnetic fields and ion densities are the highest. However, the opacity in these zones should make their contribution at VHE difficult to observe.


No $\gamma$-ray signal has been found from the SS~433/W50 interaction regions. Nevertheless, the huge jet kinetic power and the high density of  the W50 nebula, make this source one of the best microquasar/ISM interaction TeV-emitting candidates to be pursued. Our analytical treatment predicts fluxes that are slightly below the current telescopes sensitivity. Nevertheless, the improvement of the Cherenkov facilities (mostly the upcoming MAGIC and H.E.S.S. upgrades) could bring the opportunity to detect such interaction structures in the TeV regime for the first time.


\vspace{0.5cm}

\small{\textit{ACKNOWLEDGEMENTS:} The authors acknowledge support by DGI of the Spanish Ministerio de Educación
y Ci\-encia (MEC) under grant AYA2007-68034-C03-01.}


\end{document}